\documentclass[11pt,twoside]{article}
\usepackage{asp2010}

\resetcounters

\newcommand{\mlu}{\mbox{$M_{\odot}$\,yr$^{-1}$}}

\begin{document}

\title{The Global Gas and Dust budge of the Large Magellanic Cloud --- Importance of Asymptotic Giant Branch stars}
\author{M.~Matsuura$^{1,2}$
\affil{$^1$ UCL-Institute of Origins, Department of Physics and Astronomy, University College London, 
	Gower Street, London WC1E 6BT, UK}
\affil{$^2$UCL-Institute of Origins, Mullard Space Science Laboratory,
University College London, Holmbury St. Mary, Dorking, Surrey RH5 6NT, UK}}

\begin{abstract}
It is still an unresolved problem how much AGB stars can contribute to the overall gas and dust enrichment processes in the interstellar medium within galaxies. 
We start tackling this problem, by using our test case observational data from the Large Magellanic Cloud (LMC), from which we obtain the global gas and dust budget. 
The photometric data from the LMC is obtained with the Spitzer Space Telescope.
 We established an infrared colour classification scheme to select AGB stars, which are based on spectroscopically identified AGB stars. 
 We further confirm a correlation between the Spitzer colour and mass-loss rate, which leads to a measurement of the total mass-loss rate from the entire AGB population in the LMC. 
 Indeed, AGB stars are an important gas and dust source. 
 \end{abstract}

\section{The results}
It is still an unresolved problem how much AGB  stars contributes to 
gas and dust enrichment processes in the interstellar medium (ISM) within galaxies.
The Galaxy suffers a projection problem, thus it is difficult to obtain the distance to the AGB stars,
resulting in poor constraints of mass-loss rate of AGB stars.

The Spitzer Space Telescope enabled us to detect AGB stars in the neighbouring galaxy, the Large Magellanic Cloud (LMC).
 The depth of the LMC photometric survey \citep[SAGE; ][]{Meixner06} has a sufficient depth to cover
 the majority of mass-losing AGB stars \citep{Matsuura09}. The observations provided a best opportunity to study the role
 of AGB stars on chemical evolution of galaxies. \citet{Matsuura09} and Matsuura et al. (in preparation) present the details of the analysis.

 We found that indeed AGB stars are one of the important gas ejecting sources into the LMC ISM.
 Supernovae (SNe) contribute almost an equivalent amount of gas as that made from AGB stars.
 AGB stars are also the important dust source, but the relative importance remains unknown, due to the uncertainties
 in dust mass ejected from SNe.

\begin{figure}[ht]
\begin{center}
\rotatebox{270}{ 
\begin{minipage} {12cm}
\resizebox{\hsize}{!}{\includegraphics*{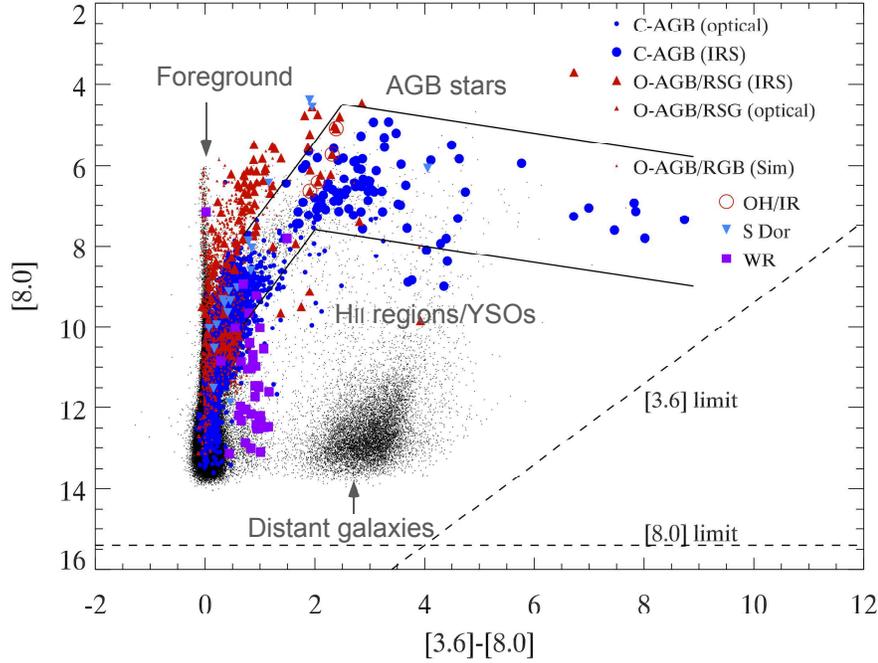}}
\end{minipage}}
 \vspace*{-1.5 cm}
 \caption{The $[3.6]-[8.0]$ vs [8.0] colour-magnitude diagram.
Spectroscopically identified oxygen-rich and
carbon-rich AGB stars, as well as S\,Dor type variable and Wolf Rayet stars  are plotted in colour symbols. 
This colour-magnitude diagram is very powerful tool for the object classification.
    \label{fig1}}
\end{center}
\end{figure}


\begin{table}[h]
  \caption{ Gas and dust mass injected from stars into the ISM of the LMC. Analysis of carbon-rich stars
   is from  \citet{Matsuura09} and that of oxygen-rich stars is from Matsuura et al. (in preparation) 
   }
 \vspace*{-0.8 cm}
\begin{center}
 \begin{tabular}{lrrlrrrrccccccccc}
  \hline
Sources & Gas mass & Dust mass & Type of dust &  \\
 & ($10^{-3}$\,\mlu)& ($10^{-6}$\,\mlu)& \\ \hline
AGB stars & 20--40$\dag$ \\
~~Carbon-rich                               &  & 40--80 & O-rich (Silicate, Al$_2$O$_3$ etc)\\
~~Oxygen-rich                               &  & 15--30 & C-rich (Amorphous carbon, MgS, graphite etc) \\
Type II SNe                                               &  20--40 & 0.1--130$\ddag$~ & both O- and C-rich \\
\hline
\end{tabular}
\label{table1}
\end{center}
 \vspace*{-0.5 cm}
$\dag$ Total of oxygen-rich and carbon-rich AGB stars
\end{table}



\end{document}